# Rational Choice Hypothesis as X-point of Utility Function and Norm Function

Takeshi Kato[1], Yasuyuki Kudo[1], Junichi Miyakoshi[1], Jun Otsuka[2], Hayato Saigo[3], Kaori Karasawa[4], Hiroyuki Yamaguchi[5] & Yasuo Deguchi[2]

[1] Hitachi Kyoto University Laboratory, Open Innovation Institute, Kyoto University, Kyoto, Japan

[2] Department of Philosophy, Graduate School of Letters, Kyoto University, Kyoto, Japan

[3] Faculty of Bioscience, Nagahama Institute of Bio-Science and Technology, Shiga, Japan

[4] Department of Social Psychology, Graduate School of Humanities and Sociology, The University of Tokyo, Tokyo, Japan

[5] Department of Behavioral and Health Sciences, Graduate School of Human-Environment Studies, Kyushu University, Fukuoka, Japan

Correspondence: Takeshi Kato, Hitachi Kyoto University Laboratory, Open Innovation Institute, Kyoto University, Kyoto 606-8501, Japan.

## Abstract

Towards the realization of a sustainable, fair and inclusive society, we proposed a novel decision-making model that incorporates social norms in a rational choice model from the standpoints of deontology and utilitarianism. We proposed a hypothesis that interprets choice of action as the X-point for individual utility function that increases with actions and social norm function that decreases with actions. This hypothesis is based on humans psychologically balancing the value of utility and norms in selecting actions. Using the hypothesis and approximation, we were able to isolate and infer utility function and norm function from real-world measurement data of actions on environmental conditions and elucidate the interaction between the both functions that led from current status to target actions. As examples of collective data that aggregate decision-making of individuals, we looked at the changes in power usage before and after the Great East Japan Earthquake and the correlation between national GDP and $CO_2$ emission in different countries. The first example showed that the perceived benefits of power (i.e., utility of power usage) was stronger than the power usage restrictions imposed by norms after the earthquake, contrary to our expectation. The second example showed that a reduction of $CO_2$ emission in each country was not related to utility derived from GDP but to norms related to $CO_2$ emission. Going forward, we will apply this new X-point model to actual social practices involving normative problems, and design the approaches for the diagnosis, prognosis and intervention of social systems by IT systems.

## 1. Introduction

In the world, various social problems are emerging e.g., disparity and inequality problems, energy problems and environmental problems. To solve these problems and aim for a sustainable, fair and inclusive society, we ought to consider social norms such as fairness, equity, morality and ethics in a broader sense. The aim of our research was to find out how to incorporate the concepts of social norms and ethics into rational decision making which was the mainstream in microeconomics, how to conceptualize and diagnose actions of individuals and groups in the real world, and how to intervene in a social system.

The major challenge facing practical approaches to solving social problems is interpreting and diagnosing individual and group behavior in terms of both "utility" and "norm". Based on the diagnosis, we should forecast the prognosis and intervene into the social system. Regarding "utility", Adam Smith said that the individual's self-interested actions leads to the unintended social benefits, as known "Invisible Hand" (Smith, 1776). Conversely, regarding "norm", Smith said that the selfish individuals also have some principles in nature to render the happiness to others (Smith, 1759). As Kenneth Arrow said, it was needed to pay attention to "social action: norms of social behavior, including ethical and moral codes" in addition to the self-interest actions (Arrow, 1970).

Based on the discussion of the Smith and Arrow, some previous researches on economic utility and social norms are known. For examples, there are the normative appropriateness in rational choice (Heath, 2008), the subjective moral costs in decision making (Basu, 2010), the community norms in social exchange game (Aoki, 2001), and the social preference in economic markets (Bowls, 2017). And, following these researches, a decision-making model is known as value function consisting of utility function of methodological individualism and norm function of social constructivism (Kato et al., 2020). The common viewpoint of these researches is that they all point out the need to incorporate social norms, that include morals and ethics, and not only economic utility, into individual decision-making models.

Utilizing these discussions on social norms, diagnosing, prognosing, and intervening into a social system, for example in social practices, requires linking real-world data to the decision-making model. That is, it is necessary to associate the actual measurement data with utility function and norm function. To date, altruism and reciprocity have been studied via experiments such as the prisoner's dilemma game and the public goods game (e.g. Bowles & Gintis, 2013), whereas empirical studies on the return potential of group norms have been conducted using questionnaire surveys (e.g. Sasaki, 2000). Game theories have tried to explain anonymous cooperative behavior by indirect reciprocity due to higher order sanctions, but they have not been observed in empirical studies (e.g. Kiyonari & Barclay, 2008). Studies of group norms have given the shape of norm function, but have not explained why individuals choosed that actions. Therefore, no inferences have been made regarding the relationship between utility and norm based on actual measurement data.

In Section 2 of this paper, we review the decision-making model from the perspectives of deontology and utilitarianism in context of previous research. We also propose a new hypothesis that reinterprets the individual's value function as the sum of utility function and norm function, and treats the choice of action as the X-point for the utility function and the norm function. In Section 3, using examples, we present the results of measurements of power usage before and after the Great East Japan Earthquake and data on national GDP per-capita and $CO_2$ emissions for different countries, isolate and infer utility function and norm function from the measurement data based on the hypothesis, and try to explain the interaction between the two functions. In Section 4, we discuss the significance of the new X-point model in understanding individual and group decision-making, as well as methods for the diagnosis, prognosis, and intervention into social systems (individual action, inter-individual interaction, and institution) by IT systems. Finally, in Section 5, we summarize our conclusions and discuss future plans towards the realization of a sustainable, fair and inclusive society.

## 2. Decision-Making Model

### 2.1 Rational Choice Model

The rational choice theory in microeconomics is one of major decision-making theories (e.g. Gilboa, 2010). It is based on the principle related to Adam Smith's Invisible Hand that individuals choose the rational action that maximizes utility based on methodological individualism. In the expected utility theory, known as the standard rational choice model, an individual assigns a certainty to beliefs in specific states, allocates a priority to desires for particular results, and maximizes the expected utility of the action. The individual utility function ($u(a)$) is expressed in equation (1). $u(o)$ is the utility of outcome ($o$), $p(o|a)$ is the probability of o given action ($a$), and $u(a)$ is the sum of multipilier of $u(o)$ and $p(o|a)$. The individual chooses the action ($a$) with the highest utility function ($u(a)$).

$$u(a) = \sum_o p(o|a)\, u(o)$$

(1)

In incorporating social norms into the rational choice model of utility function, there are two major perspectives: deontology and utilitarianism. Based on deontology, one simple method is by handling deontological constraints for actions, i.e. normative appropriateness, as same as for utilities for desires (Heath, 2008). Deontological norms are socially constructed along with language, customs, and culture. The individual's value function ($v(a)$) is expressed as the sum of utility ($u(a)$) and normative appropriateness ($n(a)$), as shown in equation (2). Here $u(a)$ is constrained to $n(a)$. For example, if $u(a)$ is an increasing function of $a$, then $n(a)$ is a decreasing function of $a$.

$$v(a) = u(a) + n(a)$$

(2)

Based on utilitarianism, other method for incorporating social norms into the rational choice model is by linking the commons game to the social exchange game (Aoki, 2001). The value function ($v(a)$) is expressed from the utility ($u_c(a)$) and the cooperation cost ($C_c(a)$) in the commons game, and the utility ($u_s(a)$) and cooperation cost ($C_s(a)$) in the social exchange game, as shown in equation (3). If the value of utility ($u_s(a)$) considering the present to the future exceeds the current cost ($C_c(a) + C_s(a)$), cooperation will be performed. Here, by regarding the utility ($u_c(a) + u_s(a)$) as $u(a)$ and the cost ($-C_c(a) - C_s(a)$) as $n(a)$, the equations (2) and (3) are innately similar on the mathematical expressions. Although the standpoints of deontology and utilitarianism are different, they can be interchanged when incorporating them as information models for social practices.

$$v(a) = u_c(a) - C_c(a) + u_s(a) - C_s(a)$$

(3)

Figure 1 shows the relationships in a rational choice model that considers social norms with the opportunity set, determinants, factors, etc. of the action. In combining equations (2) and (3), value function ($v(a)$) is expressed as the sum of the positive term ($u_p(a)$), the negative term ($u_n(a)$) of the utility function, the positive term ($n_p(a)$) and the negative term ($n_n(a)$) of the norm functions. These terms correspond to the following determinants the effectiveness and feasibility of action, economic benefit/profit and cost/risk, and ethical duty/honor and norm/obligation (e.g. Hirose, 1994) respectively. Examples of factors for each determinant are shown in Figure 1 below.

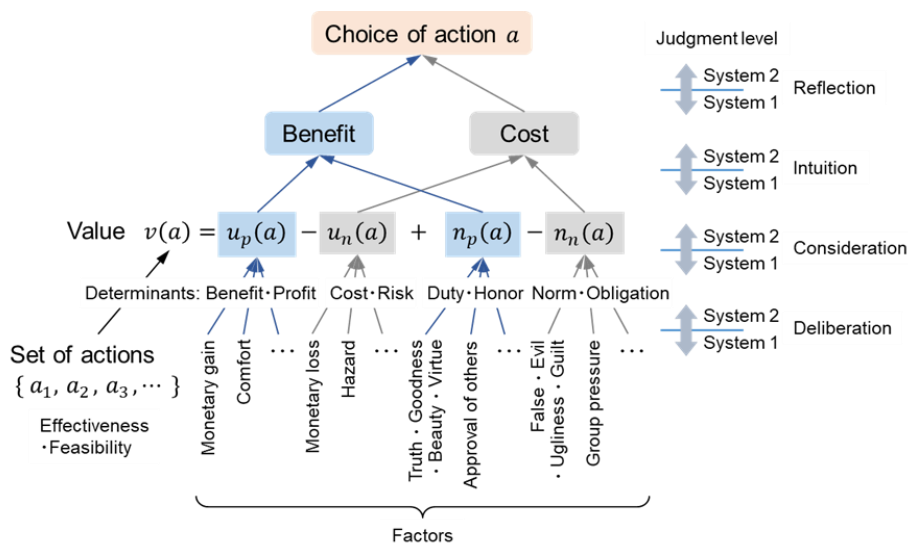

Figure 1. Choice of action that incorporates social norms

In Figure 1, determinants are associated with either utility or norm, and positive and negative terms of the utility function or norm function are aggregated as benefit and cost respectively, and ultimately, the weights of the benefit and cost are compared. The action is then chosen based on conflict. This means that the positive and negative terms of the value function intersect, and the utility and norm functions intersect. The relationship between benefit and cost shown in Figure 1 can be considered to correspond in somewhat to psychological relationships between positive and negative, individual and group, physical and social, monetary and non-monetary, direct and indirect, etc. The right side of Figure 1 shows movement of the position of the boundaries between a subjective and intuitive System 1 and an objective and reasoning System 2 (Kahneman, 2011), depending on the familiarity with or importance of the action, indicating that the level of judgment for choosing actions fluctuates between reflexive, intuitive, considerative, and deliberative.

### 2.2 X-point of Utility Function and Norm Function

To infer the relationship between utility and norm from actual measurement data in order to elucidate the reason why a certain action is chosen, first involves looking at the relationship between an action and its value. Figure 2 shows the value function ($v(x)$), utility function ($u(x)$), and norm function ($n(x)$) of an action represented as variable ($x$). Here, the diminishing utility function commonly seen in the expected utility theory is assumed as the utility function ($u(x)$), and the return potential function commonly seen in group norms is assumed as the norm function ($n(x)$). Like the relational expression of the value function ($v(a)$) in Figure 1, the value function ($v(x)$) is the sum of the utility function ($u(x)$) and the norm function ($n(x)$), and here it is assumed that action with the largest value function ($v(x)$) is the action that is chosen ($x_{vmax}$). These relationships are shown in equations (4) to (7).

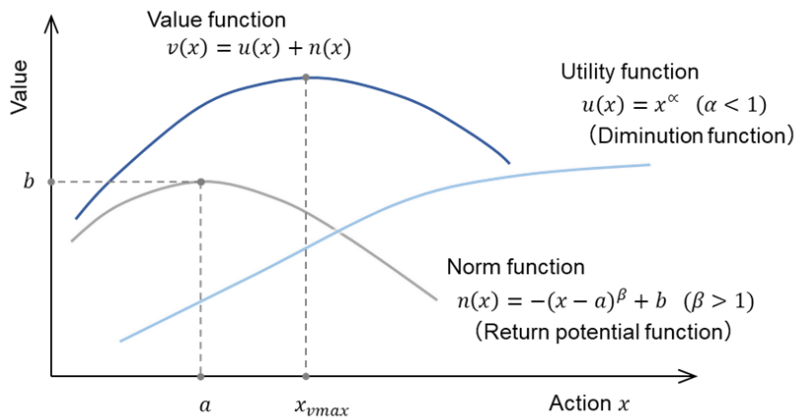

Figure 2. Relationship of action and value

$$v(x) = u(x) + n(x)$$

(4)

$$u(x) = x^\alpha \quad (\alpha < 1)$$

(5)

$$n(x) = -(x - a)^\beta + b \quad (\beta > 1)$$

(6)

$$x_{vmax} = \max_x v(x) \ \rightarrow \ \frac{dv(x)}{dx} = 0 \ \rightarrow \ \alpha \cdot x_{vmax}^{\alpha-1} - \beta \cdot (x_{vmax} - a)^{\beta-1} = 0$$

(7)

If the function forms of the utility function ($u(x)$) and the norm function ($n(x)$) are correct, measurement of the selected action ($x_{vmax}$) while changing the environmental conditions will enable identifying coefficients $\alpha$ and $\beta$, and finding the utility function ($u(x)$) and the norm function ($n(x)$). These function forms that usually take a variety of forms, however, are not always correct, thus it is difficult to find the utility function ($u(x)$) and the norm function ($n(x)$) using equation (7). There is a need, therefore, to revise the assumptions of Figure 2 and equations (4) to (7).

As a new hypothesis, therefore, we propose the interpretation of the choice of action as the X-point between the utility function ($u(x)$) and the norm function ($n(x)$). As stated in Section 2.1, actions are chosen based on the conflict between benefit and cost; i.e., the action at the X-point ($x_{veq}$) (intersection), where the utility function ($u(x)$) and the norm function ($n(x)$) are in equilibrium, becomes selected as shown in Figure 3. By assuming that linear approximation of $u(x)$ and $n(x)$ holds true within the vicinity of the X-point, the relationships between $x_{veq}$, $u(x)$, and $n(x)$ can be expressed by equations (8) to (11).

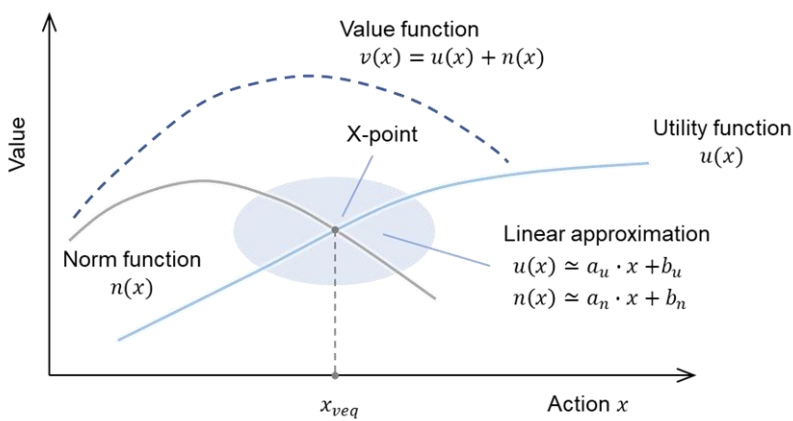

Figure 3. Choice of action as X-point of utility function and norm function

$$v(x) = u(x) + n(x) = u_b(x) - u_c(x) + n_b(x) - n_c(x)$$

(8)

$$u(x) \simeq a_u \cdot x + b_u$$

(9)

$$n(x) \simeq a_n \cdot x + b_n$$

(10)

$$u(x_{veq}) = n(x_{veq}) \;\rightarrow\; a_u \cdot x_{veq} + b_u = a_n \cdot x_{veq} + b_n \;\rightarrow\; x_{veq} = -\frac{b_u - b_n}{a_u - a_n}$$

(11)

Finding the utility function ($u(x)$) and the norm function ($n(x)$) is equivalent to finding the slope ($a_u$) and intercept ($b_u$) of $u(x)$, and the slope ($a_n$) and intercept ($b_n$) of $n(x)$. Firstly, we measure the action ($x_{veq}$) while changing the environmental conditions as shown in Figure 4 and explain the series of processes for identifying the slopes ($a_u$ and $a_n$). In equation (11) we introduce dependence on the environment ($\varepsilon$) and make the definitions as shown in equations (12) and (13).

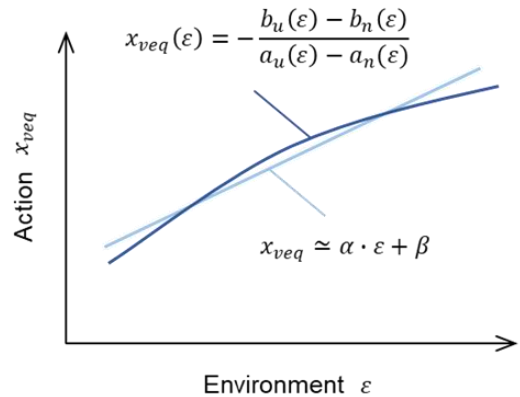

Figure 4. Dependence of action on environment

$$x_{veq}(\varepsilon) = -\frac{b_u(\varepsilon) - b_n(\varepsilon)}{a_u(\varepsilon) - a_n(\varepsilon)} = -\frac{(\mu_u - \mu_n) \cdot \varepsilon + (\nu_u - \nu_n)}{(\kappa_u - \kappa_n) \cdot \varepsilon + (\lambda_u - \lambda_n)}$$

(12)

$$a_u \simeq \kappa_u \cdot \varepsilon + \lambda_u$$
$$a_n \simeq \kappa_n \cdot \varepsilon + \lambda_n$$
$$b_u \simeq \mu_u \cdot \varepsilon + \nu_u$$
$$b_n \simeq \mu_n \cdot \varepsilon + \nu_n$$

(13)

Next, by fitting the results of measurements made while changing the environment ($\varepsilon$) into equation (12), we find the four coefficients ($\kappa$, $\lambda$, $\mu$, $\nu$) shown in the following equations (14).

$$\kappa = \kappa_u - \kappa_n$$
$$\lambda = \lambda_u - \lambda_n$$
$$\mu = \mu_u - \mu_n$$
$$\nu = \nu_u - \nu_n$$

(14)

Moreover, in order to solve the equations in (14), the following equations (15) are given as two constraint conditions, for example, to divide $\kappa$ into $\kappa_u$ and $\kappa_n$. One is the environmental condition ($\varepsilon_{uo}$) by which the slope ($a_u$) of the utility function ($u(x)$) becomes 0, and the other is the environmental condition ($\varepsilon_{no}$) by which the slope ($a_n$) of the norm function ($n(x)$) becomes 0. The former means that utility cannot be gained by performing actions above or below the condition, and the latter means that performing actions below or above the condition will eliminate the normative constraints.

$$a_u \simeq \kappa_u \cdot \varepsilon_{uo} + \lambda_u = 0$$
$$a_n \simeq \kappa_n \cdot \varepsilon_{no} + \lambda_n = 0$$

$$(15)$$

Thus, by solving equations (14) and (15) as a system of equations, we can find coefficients $\kappa_u$, $\lambda_u$, $\kappa_n$ and $\lambda_n$ as shown by the equations in (16), and identify the slopes ($a_u$ and $a_n$) by incorporating these coefficients in equations in (13).

$$\kappa_u = \frac{\kappa \cdot \varepsilon_{no} + \lambda}{\varepsilon_{no} - \varepsilon_{uo}} \qquad \lambda_u = -\varepsilon_{uo} \cdot \frac{\kappa \cdot \varepsilon_{no} + \lambda}{\varepsilon_{no} - \varepsilon_{uo}}$$

$$\kappa_n = \frac{\kappa \cdot \varepsilon_{uo} + \lambda}{\varepsilon_{no} - \varepsilon_{uo}} \qquad \lambda_n = -\varepsilon_{no} \cdot \frac{\kappa \cdot \varepsilon_{uo} + \lambda}{\varepsilon_{no} - \varepsilon_{uo}}$$

$$(16)$$

After finding the slopes ($a_u$ and $a_n$), we need to find the intercepts ($b_u$ and $b_n$). We assume that a linear approximation holds true for the results of measurement of action ($x_{veq}$) in response to the environment ($\varepsilon$), as shown in equation (17) and in Figure 4. By matching equation (17) with equations (12) and (14) we then yield equations (18).

$$x_{veq} \simeq \alpha \cdot \varepsilon + \beta$$

$$(17)$$

$$\mu = -\alpha \cdot \lambda$$
$$\nu = -\beta \cdot \lambda$$
$$\kappa = 0$$
$$\lambda\, (= 1)$$

$$(18)$$

We substitute equations (18) with equations (16) to obtain equations (19), and then find the utility function ($u(x)$) and the norm function ($n(x)$), as shown from equations (13) and (14) in (20).

$$\kappa_u = \frac{\lambda}{\varepsilon_{no} - \varepsilon_{uo}} \qquad \lambda_u = -\varepsilon_{uo} \cdot \frac{\lambda}{\varepsilon_{no} - \varepsilon_{uo}}$$

$$\kappa_n = \frac{\lambda}{\varepsilon_{no} - \varepsilon_{uo}} \qquad \lambda_n = -\varepsilon_{no} \cdot \frac{\lambda}{\varepsilon_{no} - \varepsilon_{uo}}$$

$$(19)$$

$$u(x) \simeq a_u \cdot x + b_u = \lambda \cdot \frac{\varepsilon - \varepsilon_{uo}}{\varepsilon_{no} - \varepsilon_{uo}} \cdot x - (\lambda \cdot \alpha - \mu_n) \cdot \varepsilon - (\lambda \cdot \beta - \nu_n)$$

$$n(x) \simeq a_n \cdot x + b_n = \lambda \cdot \frac{\varepsilon - \varepsilon_{no}}{\varepsilon_{no} - \varepsilon_{uo}} \cdot x + (\lambda \cdot \alpha + \mu_u) \cdot \varepsilon + (\lambda \cdot \beta + \nu_u)$$

$$(20)$$

Here, since the intercepts ($b_u$ and $b_n$) are positioned relative to the X-point between the utility function ($u(x)$) and the norm function ($n(x)$), by moving the origin of equations (20), we can obtain equations (21).

$$u'(x) \simeq \lambda \cdot \frac{\varepsilon - \varepsilon_{uo}}{\varepsilon_{no} - \varepsilon_{uo}} \cdot x$$

$$n'(x) \simeq \lambda \cdot \frac{\varepsilon - \varepsilon_{no}}{\varepsilon_{no} - \varepsilon_{uo}} \cdot x + (\lambda \cdot \alpha + \mu_u) \cdot \varepsilon + (\lambda \cdot \beta + \nu_u) + (\lambda \cdot \alpha - \mu_n) \cdot \varepsilon + (\lambda \cdot \beta - \nu_n)$$

$$\simeq \lambda \cdot \frac{\varepsilon - \varepsilon_{no}}{\varepsilon_{no} - \varepsilon_{uo}} \cdot x + 2 \cdot \lambda \cdot \alpha \cdot \varepsilon + (\mu_u - \mu_n) \cdot \varepsilon + 2 \cdot \lambda \cdot \beta + (\nu_u - \nu_n)$$

$$\simeq \lambda \cdot \frac{\varepsilon - \varepsilon_{no}}{\varepsilon_{no} - \varepsilon_{uo}} \cdot x + \lambda \cdot \alpha \cdot \varepsilon + \lambda \cdot \beta$$

$$(21)$$

Equations (21) shows that the action ($x_{veq}$) can be interpreted as the X-point between the utility function ($u'(x)$) and norm function ($n'(x)$), which are isolated from the two environmental conditions ($\varepsilon_{uo}$, $\varepsilon_{no}$) and from the constants $\alpha$ and $\beta$ obtained from the results of measurement of action ($x_{veq}$) in response to the environment ($\varepsilon$).

In Figure 5, we would like to suppress the current relationship of action and environment ($x_{veq} \simeq \alpha \cdot \varepsilon + \beta$) into the target relationship of action and environment ($x'_{veq} \simeq \alpha' \cdot \varepsilon + \beta'$).

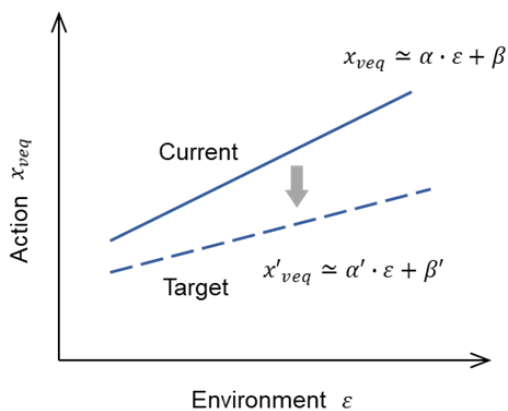

Figure 5. Current status and target of action on environment

Assuming that the utility function ($u'(x)$) and norm function ($n'(x)$) can be obtained from the current results of measurement of action ($x_{veq}$) in response to the environment ($\varepsilon$) as shown in Figure 6, there are four approaches (I, II, III and IV) to achieving the target, depending on the slope ($a_u$) and intercept ($b_u$) of $u'(x)$, and the slope ($a_n$) and intercept ($b_n$) of $n'(x)$, as shown in Table 1. Among these, approaches II and IV, which pertain to norms, correspond respectively to "marginal crowding in" and "categorical crowding in" mentioned by Bowles may be useful in considering the effects of economic incentives on norms (Bowls, 2017).

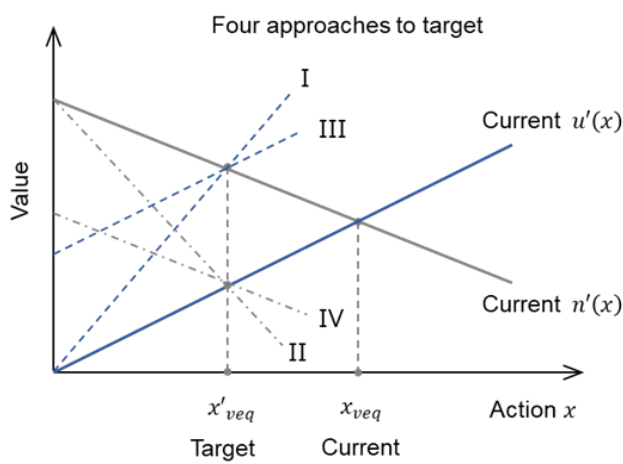

Figure 6. Four approaches from current to target

Table 1. Details for four approaches from current to target

| Approach | Detail | |
|---|---|---|
| I | Increase the positive slope $a_u$ of $u'(x)$ (sensitivity to action)<br>⇒ Increase $\varepsilon_{uo}$ | $a_u \propto \dfrac{\varepsilon - \varepsilon_{uo}}{\varepsilon_{no} - \varepsilon_{uo}}$ |
| II | Decrease the negative slope $a_n$ of $n'(x)$ (sensitivity to action)<br>⇒ Increase $\varepsilon_{no}$ | $a_n \propto -\dfrac{\varepsilon_{no} - \varepsilon}{\varepsilon_{no} - \varepsilon_{uo}}$ |
| III | Increase the intercept $b_u$ of $u'(x)$ (base of action)<br>⇒ Raise the level of utility | |
| IV | Decrease the intercept $b_n$ of $n'(x)$ (base of action)<br>⇒ Suppression by norm | |

## 3. Examples

### 3.1 Power Usage

To test the new hypothesis on rational choice as the X-point of utility function and norm function, we applied it to power usage measurements in response to outdoor temperature within areas serviced by Tokyo Electric Power Company before (August 2010) and after (August 2011) the Great East Japan Earthquake (Chikamoto, 2012). Temperatures were amended to incorporate the effect of temperatures in the preceding two days. Although this example does not involve decision-making by individuals, it illustrates changes in group decision-making and group norms before and after a disaster as an aggregate measure of individual decision-making. From the results shown in Figure 7, we used equations (22) and equations (23) to represent the correlation between power usage and temperature before the disaster and after the disaster respectively. In these equations, temperature corresponds to environment ($\varepsilon$) and power usage corresponds to action ($x_{veq}$). The constraint conditions for the environment and the action were derived based on Figure 7. The temperatures at which the residents started to use power were set as the $\varepsilon_{uo}$ before and after the disaster. The $\varepsilon_{no}$ was set to 30 ℃ since the slight inflection point seen from the plot before and after the disaster indicates that normative constraints lost their effect once the temperature exceeded 30 ℃ due to the residents' health.

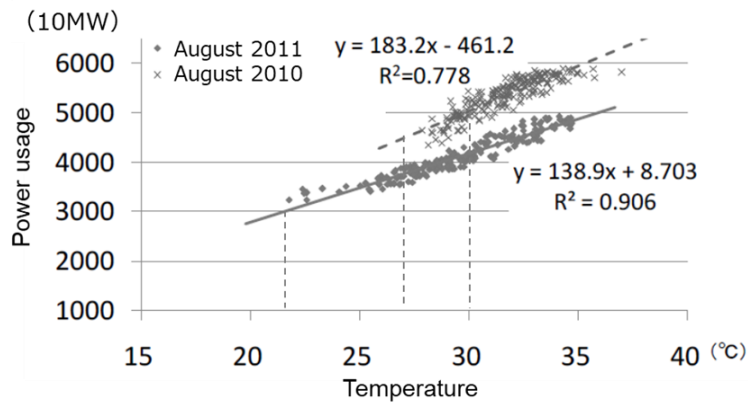

Figure 7. Power usage in areas serviced by Tokyo Electric Power Company before and after the Great East Japan Earthquake Disaster (Chikamoto, 2012. Reprinted by permission of the author)

Before:
$$x_{veq} \simeq 183.2 \cdot \varepsilon - 461.2 \qquad \varepsilon_{uo} = 27\ ℃ \qquad \varepsilon_{no} = 30\ ℃$$

(22)

After:
$$x_{veq} \simeq 138.9 \cdot \varepsilon + 8.703 \qquad \varepsilon_{uo} = 22\ ℃ \qquad \varepsilon_{no} = 30\ ℃$$

(23)

Solving equations (22) and (23) (by answering equations (17) to (21) shown in Section 2) yielded the equations (24) and

(25) concerning utility function ($u'(x)$) and norm function ($n'(x)$) for the measurement results. Figure 8 shows the relationship between action $x$ and $u'(x)$, $n'(x)$ before the disaster, and Figure 9 shows their relationship after the disaster. In Figure 8 and Figure 9, with the outdoor temperature as a parameter, the lines rising to the right show the utility functions (blue lines) and the lines rising to the left show the norm functions (gray lines), and the X-point of the two functions represents the choice point of the action.

Before:

$$u'(x) \simeq \frac{\varepsilon - 27}{3} \cdot x \qquad n'(x) \simeq -\frac{30 - \varepsilon}{3} \cdot x + 183.2 \cdot \varepsilon - 461.2$$

$$(24)$$

After:

$$u'(x) \simeq \frac{\varepsilon - 27}{3} \cdot x \qquad n'(x) \simeq -\frac{30 - \varepsilon}{3} \cdot x + 183.2 \cdot \varepsilon - 461.2$$

$$(25)$$

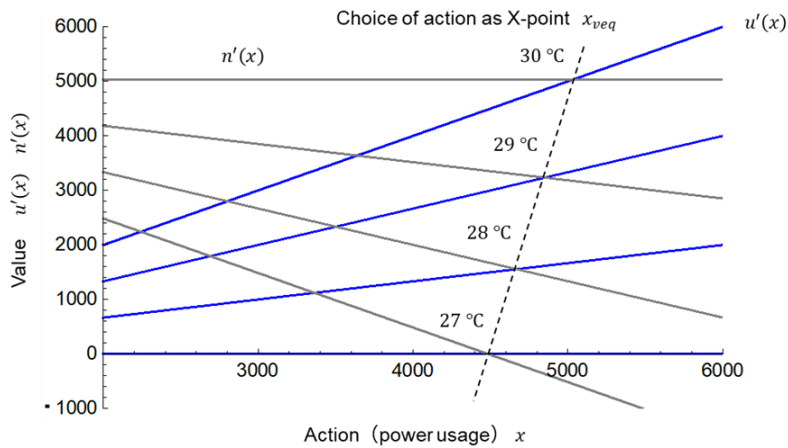

Figure 8. Utility functions and norm functions before the disaster

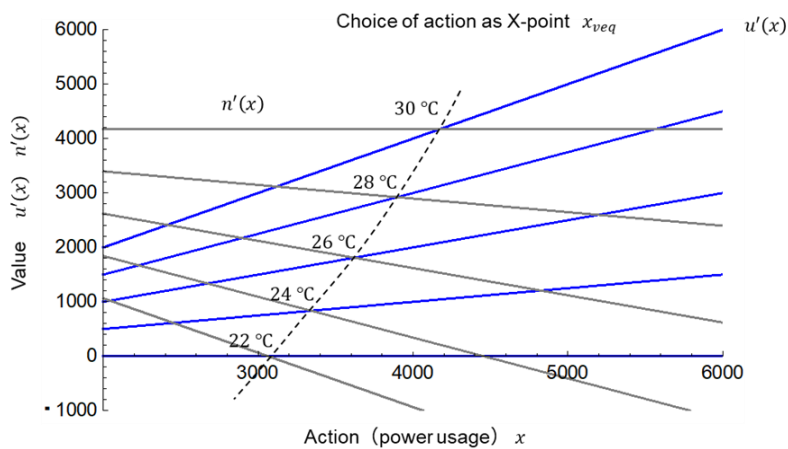

Figure 9. Utility functions and norm functions after the disaster

Figure 10 shows a comparison of utility function and norm function before and after the disaster when the environmental temperature ($\varepsilon$), was 28℃. The slope of the utility function ($u'(x)$) was larger after the disaster compared to that before the disaster, which is consistent with approach I shown in Table 1. In other words, the utility gained from using power increased after the disaster. Unexpectedly, for the norm function ($n'(x)$), the slope was smaller after the disaster compared to that before the disaster, which is the opposite of approach II shown in Table 1. In other words, after the disaster, rather than simply and strictly controlling their power usage, residents tried to use power more efficiently and carefully, based on the relationship of utility and norms. This implied that although residents started using power at a lower temperature ($\varepsilon_{uo}$) after the disaster, they in fact reduced their overall power usage.

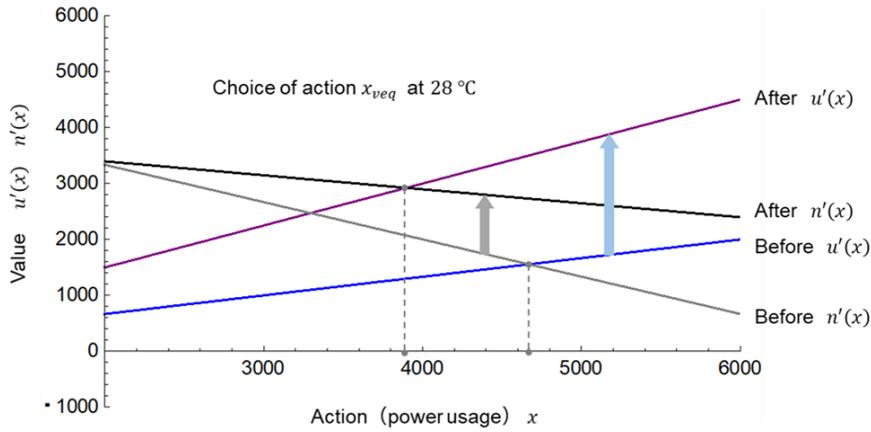

Figure 10. Comparison of utility functions and norm functions before and after the disaster

### 3.2 CO₂ Emissions

Another example based on actual measurements was drawn from data on the relationship between CO$_2$ emissions and to gross domestic product (GDP) per capita in different countries around the world (Archer, 2018). The country is not an individual decision-maker but a collection of individual decision-makers. Figure 11 is a plot of the GDP per capita (current US$) against CO$_2$ emissions (metric tons per capita) for 2013 based on open data from the World Bank (The World Bank). By analyzing the plotted data, we used equations (26) and (27) to represent the correlation between GDP and CO$_2$ emissions, separately for countries with lower CO$_2$ emissions (mainly developed nations) and countries with higher CO$_2$ emissions (mainly developing nations). In these equations, economic environment expressed by GDP represents the environment ($\varepsilon$), and CO$_2$ emission represents the action ($x_{veq}$). As regards the constraint condition, since norms become meaningless when both GDP and CO$_2$ emission are 0, we set the constraint condition as $\varepsilon_{no} = 0$ \$. Additionally, on the basis of a study on GDP and life satisfaction (Eurostat, 2018), we assumed that utility reaches saturation at $\varepsilon_{uo}$ of 30,000 \$ or higher.

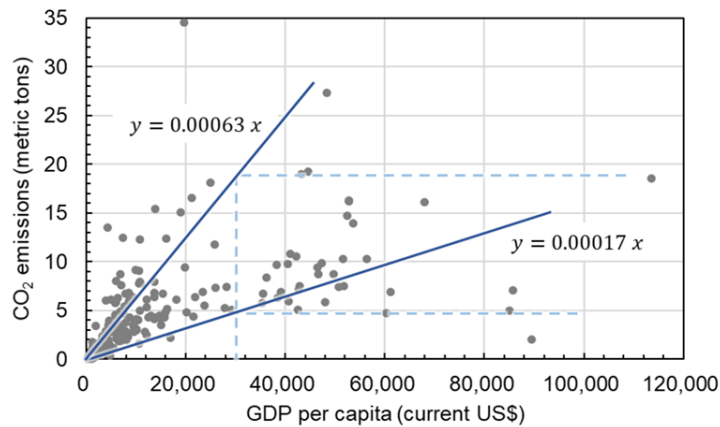

Figure 11. GDP per capita and CO₂ emissions (Archer, 2018; The World Bank)

Countries with low CO$_2$ emissions:

$$x_{veq} \simeq 0.00017 \cdot \varepsilon \qquad \varepsilon_{uo} = 30{,}000 \ \$ \qquad \varepsilon_{no} = 0 \ \$$$

(26)

Countries with high CO$_2$ emissions:

$$x_{veq} \simeq 0.00063 \cdot \varepsilon \qquad \varepsilon_{uo} = 30{,}000 \ \$ \qquad \varepsilon_{no} = 0 \ \$$$

(27)

Solving equations (26) for countries with low CO$_2$ emissions and equations (27) for countries with high CO$_2$ emissions yielded equations (28) and (29) respectively for their utility function ($u'(x)$) and norm function ($n'(x)$). Figure 12 shows a comparison of countries with low and those with high $CO_2$ emissions when GDP, representing the environment ($\varepsilon$), is at 20,000 \$. Although the utility functions ($u'(x)$) of the two groups are the same, the intercepts of their norm functions ($n'(x)$) differ, which is consistent with approach IV shown in Table 1. Although the utility provided by economic activities that result in the release of $CO_2$ is the same between the countries with low $CO_2$ emissions and those with high $CO_2$ emissions, the bases of the norms that serve to suppress the $CO_2$ emissions for the two groups of countries are different. As shown in the COP (the conference of the parties to the United Nations Framework Convention on Climate Change), it can be said that reducing $CO_2$ emissions on a global scale requires nurturing the norms particular to each country, and institutionalizing reduction measures based on those norms. Reducing $CO_2$ emissions, therefore, is evidently related to norms.

Countries with low $CO_2$ emissions:

$$u'(x) \simeq \frac{30000 - \varepsilon}{30000} \cdot x \qquad n'(x) \simeq -\frac{\varepsilon}{30000} \cdot x + 0.00017 \cdot \varepsilon$$

(28)

Countries with high $CO_2$ emissions:

$$u'(x) \simeq \frac{30000 - \varepsilon}{30000} \cdot x \qquad n'(x) \simeq -\frac{\varepsilon}{30000} \cdot x + 0.00063 \cdot \varepsilon$$

(29)

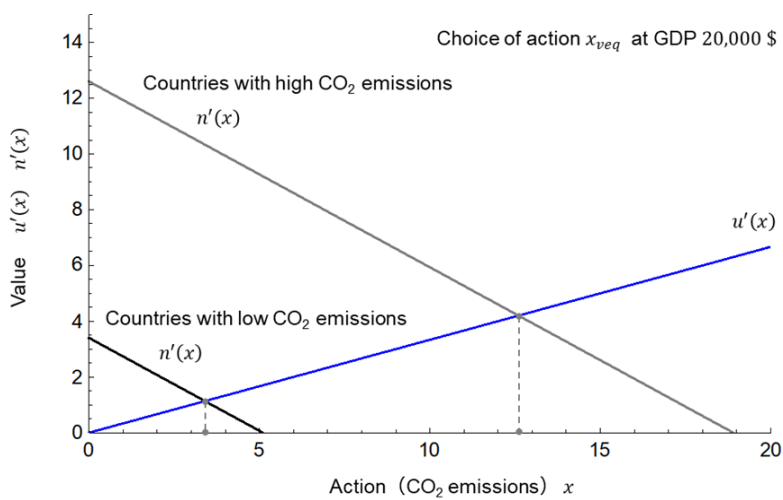

Figure 12. Comparison of countries with high and low $CO_2$ emissions

## 4. Discussions

### 4.1 Review of Results

Through the above two examples, we were able to isolate utility function and norm function from actual measurement data and understand their interaction by applying the hypothesis that rational choices are made based on the X-point between utility function and norm function. In the example pertaining to power usage before and after disaster, rather than enforcing stricter norms, using power with a sense of appreciation is more effective in suppressing power usage. This corresponds to the results of an internet-based survey conducted after the earthquake on household electricity saving behavior that showed that mainly perceived benefits and not social norms significantly affected power usage (Yagita, Iwafune, Ogiwara & Fujimoto, 2012). In the example pertaining to GDP and $CO_2$ emission by different countries, there were no differences in the utility provided by economic activities that resulted in the release of $CO_2$, and the norms of each country played a significant role in reducing $CO_2$ emissions. These conclusions can be considered as new knowledge gained by applying our hypothesis.

This theory is based on the principle illustrated in Figure 1, where humans are said to perform a certain level of that balances the benefits and costs in choosing actions. This decision-making process is not only based on monetary value, such as profits and costs, but also involves assigning certain weights to non-monetary value, such as responsibilities and norms and comparing the trade-offs between monetary and non-monetary values. It is consistent with the fork psychology that we apply in everyday life. Additionally, the existence of various ways for selecting actions, such as making reflexive decisions or deciding after thorough deliberation, can be understood as resulting from the movement of the boundary between System 1 and System 2, depending on the familiarity with or importance of the action. In particular, changing this decision level is important in evoking and nurturing norms.

We however, were not able to investigate the problem of inseparability between the moral sentiments or social preferences and the economic incentives, as mentioned by Bowles, in the context of the conventional real world (Bowls, 2017). Our hypothesis, however, allowed us to interpret the interaction between utility function and norm function. We demonstrated through our hypothesis that when economic incentives crowd out the moral sentiments, the "marginal" slope or the "categorical" intercept of the norm function decreases inversely to the slope or the intercept of the utility function. "Crowding out" in economics originally refers to how interest rates increase and suppress the economic activities of citizens, despite the implementation of economic measures through the issue of public bonds. Bowles uses the term "crowding out" to refer to the suppression of moral sentiments by incentives. On the other hand, when the incentives crowd in the moral sentiments, demonstrating that the norm function increases following the increase in utility function and distinguishing between the two functions has significant implications in the proposal and design of public policies.

*4.2 Effects*

So far, the interpretation of real-world actions has been investigated in terms of the assignment of meaning to actions based on the function form of the utility function in relation to the rational choice theory, and through experiments on altruism and reciprocity in relation to the game theory and surveys on group norms by questionnaire. We believe that the isolation and visualization of utility function and norm function from actual measurement data through our hypothesis will drive the progress of conventional research and is vital in promoting social practices. Diagnosis and prognosis based on our hypothesis will be important in assessing real-time normative interventions in social systems.

In our hypothesis, we used constraint conditions and dependence on the environment to find the utility function and the norm function. If constraint conditions and dependence of the action on parameters other than the environment were known, they may also be used to find the utility function and the norm function using the same procedure. The important contribution of our hypothesis lies in being able to connect actual measurement data to utility function and norm function.

Although the examples presented in this study pertain to groups of people serviced by the Tokyo Electric Power Company and citizens of countries of the world, our hypothesis can also be applied to actions of individuals. Therefore, it can be useful in understanding personality and characteristics of groups as well as individuals. The relationship between choice of actions and the rational choice model that incorporates social norms shown in Figure 1 and the determinants of the action and the factors related to the determinants are related to behavioral economics and the nudge theory (method for inducing behavior). Individual diagnosis on utility and norms based on our hypothesis can be used in urging individuals to take normative actions.

In envisaging a future of the social system, for example, the Japanese government proposes "Society 5.0." Society 5.0 is defined as "a human-centered society that balances economic advancement with the resolution of social problems by a system that highly integrates cyberspace and physical space" (Cabinet Office, Government of Japan, 2014). Until the information society (Society 4.0), the analysis of data and the explanation and prediction of phenomena were performed according to astronomical and physical paradigms. In Society 5.0, the integration of cyberspace (IT systems) and physical space (social systems) has enabled the diagnosis and prognosis of social phenomena in real time according to clinical medical paradigms, leading to the clinical intervention of social systems by IT systems (Deguchi, Otsuka, Kudo & Kato, 2018a).

Combining the human-centric perspective with the perspective of integrating society and IT in Society 5.0, as shown in Figure 13, entails IT systems to perform diagnosis and prognosis of social systems and carry out real-time normative and ethical interventions to social systems based on the diagnosis and prognosis (Deguchi et al., 2018a; Deguchi, Kato, Kudo, Karasawa & Saigo, 2018b; Karasawa et al., 2018; Kato et al., 2020). In addition to targeting social systems composed of groups of people, IT systems must comprehensively capture the macro-level interactions between individuals as well as the meta-level social institutions shared by the groups, based on the micro-level decision-making activities of individuals. In other words, individual decision-making must take into consideration the influence of inter-individual interaction and social institutions, and the effect of social constructivistic normativity, in addition to individualistic utility. Diagnosis of utility and norm functions based on our hypothesis can help in intervening at micro, macro and meta levels by IT systems.

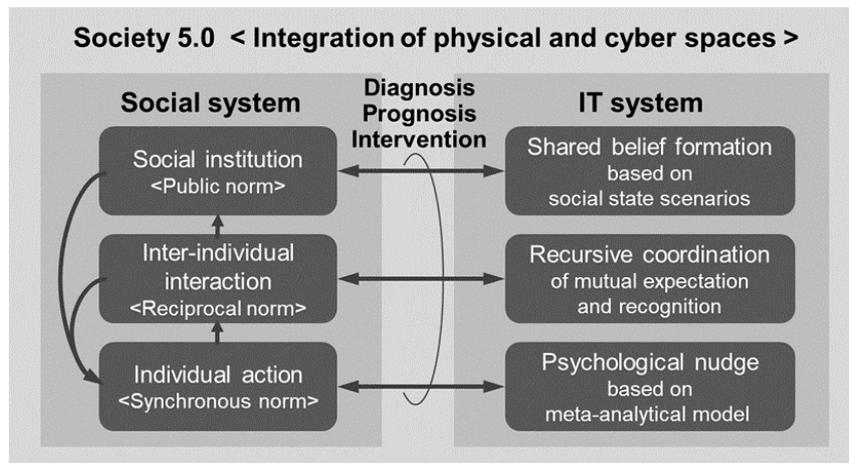

Figure 13. Normative intervention of social systems by IT systems in Society 5.0 (Kato et. al., 2020)

## 5. Summary

### 5.1 Conclusions

To attain a sustainable, fair and inclusive society, we proposed a new decision-making model that incorporated social norms into a rational choice model. On this basis that interpreted a selected action as the X-point between utility function and norm function, we then presented a method for isolating and inferring utility function and norm function from actual measurement data, and interpreted the interaction between the two functions using actual examples.

1) In place of a standard rational choice model based on methodological individualism, we presented a rational choice model that incorporated social norms from the standpoints of both deontology and utilitarianism. This model expressed the value function in decision-making as the sum of the utility function and norm function. The opportunity set of actions, the determinants, and the various factors are connected together for each of the functions, their positive and negative terms are then consolidated respectively into the weights for benefit and cost, and the action is eventually chosen by comparing the weights.

2) To distinguish the utility function and norm function from real-world measurement data, we presented a new hypothesis that interprets the choice of action as the X-point between the utility function and the norm function. Using this hypothesis, we derived the relational expressions for finding the utility function ($u'(x)$) and the norm function ($n'(x)$) from the linear approximation equation ($\alpha \cdot \varepsilon + \beta$) of the measurement data for the action ($x_{veq}$) in response to the environment ($\varepsilon$), and the two constraint conditions ($\varepsilon_{uo}$, $\varepsilon_{no}$) related to utility and norm. We also presented four approaches to move the current action closer to the target action, based on the slopes and intercepts of the two functions obtained from the measurement data.

3) To assess the new hypothesis, we used it to analyze two examples as aggregates of individual decision making, namely the measurements of power usage before and after the Great East Japan Earthquake and data on national GDP and $CO_2$ emissions. The former example showed that instead of restriction of power use due to stricter norms, residents were trying to effectively use power based on appreciation of its benefits. In the latter example, comparison of countries with low and with high $CO_2$ emissions showed that they differed not only in terms of the utility of the economic activities, but also in terms of the basis of the norms that restricted $CO_2$ emission.

4) After examining the hypothesis through the above two examples, we considered that the new hypothesis will lead to new interpretations of real-world measurement data, and understanding of psychological balancing of benefits and costs in making daily choices of action. It can also contribute to "crowding in" of norms by economic incentives and in designing public policies, and can be applied in the diagnosis, prognosis, and intervention of social systems (individual action, inter-individual interaction, and institution) by IT systems.

### 5.2 Future Prospects

To verify our proposed hypothesis, we will collect examples for commons games, trust games, and other game experiments, as well as for natural experiments, and demonstrate the effectiveness of our hypothesis by comparing interpretations based on the hypothesis with conventional interpretations. Concurrently, we will conduct field trials of normative interventions by IT systems using the hypothesis through experiments on cycling of renewable energy in the context of the regional economy.

1) Individual action: Examine whether persuasion using nudge or provision of information affects utility or norm, and whether they can incite normative action, by isolating the utility function and norm function from actual measurement data for individuals and families.

2) Inter-individual interaction: Examine how expectation and recognition from others affect utility function and norm function for individuals and families, and whether they would change the results of optimization of distributed cooperative control using IT systems.

3) Social institutions: Find the utility function and norm function from actual measurement data for groups composed of individuals and families and consider what kind of intervention should be done to social institutions represented by the group's common expectations.